# Word Existence Algorithm

Tejeswini Sundaram
Dept. of Computer Science & Engineering,
Manipal Institute of Technology,
Manipal – 576104, India
tejeswinisundaram@live.com

Vyom Chabbra
Dept. of Computer Science & Engineering,
Manipal Institute of Technology,
Manipal – 576104, India.
vyomchabbra@live.com

*Abstract*— the current scenario in the field of computing is largely affected by the speed at which data can be accessed and recalled. In this paper, we present the word existence algorithm which is used to check if the word given as an input is part of a particular database or not. We have taken the English language as an example here. This algorithm tries to solve the problem of lookup by using a uniformly distributed hash function. We have also addressed the problem of clustering and collision. A further contribution is that we follow a direct hashed model where each hash value is linked to another table if the continuity for the function holds true. The core of the algorithm lies in the data model being used during preordering. Our focus lies on the formation of a continuity series and validating the words that exists in the database. This algorithm can be used in applications where we there is a requirement to search for just the existence of a word, example Artificial Intelligence responding to input ,look up for neural networks and dictionary lookups and more. We have observed that this algorithm provides a faster search time

## I. INTRODUCTION

Development in computer software and applications continues to be very dynamic. Each software problem requires different tools and algorithms to solve it effectively and achieve best results. As a result, we witness the announcement of new tools and algorithms in quick succession with multiple updates.

Among these algorithms, there have been various solutions being provided for solving the look up or word search problem. It is interesting to note that, though there seems to be multiple choices for looking up an entity, there is not one solution which provides optimal results in terms of both time and space complexity. The algorithms are selected depending on the nearest problem we require to be solved.

With the help of the currently available software tools, there are a multitude of desktop and mobile applications that are being developed on various platforms, be it android, IOS or windows phone application. Mobile application development has sprung over all industries and domains, from e-commerce to food, from health to education, from gaming to fashion, app development has taken a huge leap. Every aspect of human life has been effected by mobile technology and hence there is a requirement for building much faster responsive applications.

Many of these mobile applications, require a fast and simple search algorithm to perform their tasks. These applications are generally light weight in nature and require simple look-up algorithms that are not time-consuming to serve their purpose. We can observe that the biggest pain point for an application is the experience of waiting for a page to respond or load. There are various reasons for slow response of applications. Look-up or searching for an entity in the database is one of the reasons causing this delay.

Other areas where look-up are required is in the development of artificial neural networks and the initial stages of an artificial intelligence that respond to input, and dictionary look-ups. It has been noticed that the major requirement for all the above applications is fast search time. The traditional hash function provide a considerable solution to this problem, but still has certain problems like collision and clustering.

In this paper, we attempt to provide a simple, yet effective hash function to solve the problem of look-up for a word in any database or table data.

## II. METHODOLOGY

The search defines a specific storage pattern to work properly. Once the data is stored in the proper format we can check if a particular word exists in the list or not. Adding of new values to this storage type is relatively simple. Data is stored in form of nested hash tables.

For this explanation: we take around 50000 words in the English language to be stored (26 possible characters).The length of a word (n) specifies the no. of levels (l) we need to store it. Each level will have maximum $26^n$ hash tables in it where n is the length of word.

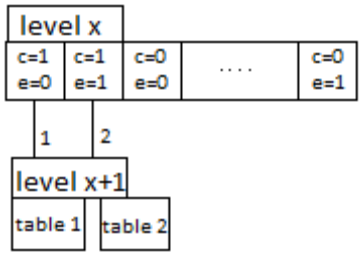

Fig 1. Levels in Hash Table

Each table has 26 inputs with two values stored in it, the first is for the continuation value and the second one is for the existence value. These inputs are then connected to a table of their own on the next level depending on the continuation value. For example take into consideration this case when for an input when the continuation value is zero. This invariably means that no word with such a prefix exists. Thus we can stop the search here. On the other hand, if c = 1 then we can keep searching and move to the next level in the Hash table.

Once we reach the last character of the searched word, we then check for the existence value instead of the continuation value. If the existence value is 1(true) then the algorithm returns that the word is valid, otherwise it is invalid

### III. EXPERIMENTAL RESULTS

The above algorithm is advantageous in the terms that it works very fast with minimal time complexity and exhibits space overwriting.

For explanation purposes, consider two search words "*bat*" and "*bath*" to be searched in the hash table. The format of the levels of the hash table is as shown in Figure 2. Note that the letter 'c' depicts the continuation value and 'e' depicts the existence value as discussed in the earlier paragraph.

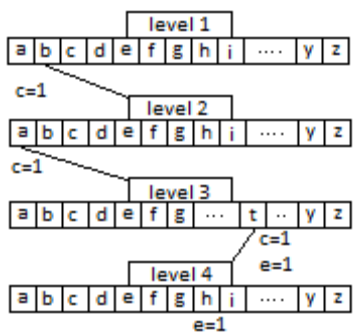

Here we can observe that the two words '*bat*' and '*bath*' are stored in 4 levels. And 3 of the layers are shared by both.

This sharing of spaces can done for any number of combinations and thus can reduce the total space needed for storage.

The Word-Existence Algorithm discussed in the previous section is a based on a simple hash function. It inherits the features of a uniformly distributed hash function and is an example of minimal perfect hashing. The problems of clustering is eliminated by the multi-level hash table model that has been incorporated. No two characters are stored in the same bin of the hash table. A separate bin is allotted to all the 26 characters in the English alphabets. Hence, clustering and collision has been avoided. Another important feature of the Word Existence Algorithm is that it allocates a bin (memory) to the alphabet, only if it is necessary. This way, memory space has been saved and unnecessary allocations will be avoided.

Here, we follow a direct hashed model where each hash value is linked to another table if the continuity for the function holds true. The data model used during pre-ordering is the highlight of the algorithm. We have observed that this multi-hash algorithm provides a faster search time than the conventional searching techniques with respect to the applications discussed earlier and the time complexity is found to be of the order O(1).

### IV. CONCLUSION

This algorithm was developed keeping in mind the application of it in database retrieval and management, mobile apps, desktop apps and various other projects that require basic look-up tasks. The algorithm tries to solve the problem of waiting for a page to load or respond to an event that has occurred. Look-up in the database is one of the reasons causing the delay in loading, and by speeding up the look-up we have intended to speed up the loading process. Our experimental results show that, the algorithm is well suited for applications that require fast and simple look-up from an existing database. The highlights of the algorithm is that it is a minimally perfect hash function which has efficient memory management for simple databases. It is based on multi-level hashing and is very quick for the discussed applications.